\documentclass[12pt]{article}
\usepackage{amssymb,float}
\usepackage{amsmath}
\usepackage{graphicx}
\usepackage[T1]{fontenc}
\usepackage{lmodern}
\usepackage{textcomp}
\usepackage{bm}
\newcommand{\mathsym}[1]{{}}

\usepackage{graphicx}
\usepackage{rotating}
\usepackage{color}
\newcommand{\bra}{\begin{array}}
	\newcommand{\era}{\end{array}}
\newcommand{\beq}{\begin{equation}}
\newcommand{\eeq}{\end{equation}}
\newcommand{\beqar}{\begin{eqnarray}}
\newcommand{\eeqar}{\end{eqnarray}}

\newcommand{\be}{\begin{equation}}
\newcommand{\ee}{\end{equation}}
\newcommand{\bea}{\begin{eqnarray}}
\newcommand{\eea}{\end{eqnarray}}
\newcommand{\bd}{\begin{displaymath}}
\newcommand{\ed}{\end{displaymath}}

\newcommand{\h}{\hbar}

\newcommand{\al }{\alpha}

\begin{document}
	
	\vspace{20pt}
	
	\begin{center}
		
		{\LARGE \bf Investigation of Unruh temperature of Black holes by using of EGUP formalism

			\medskip
		}
		\vspace{15pt}
	{\large Hassan Hassanabadi${}^{1,4}$,  Nasrin Farahani ${}^{1,\dag}$, Won Sang Chung${}^{2}$, and\\ Bekir Can L\" utf\"uo\u{g}lu${}^{3,4}$}
		
		\vspace{15pt}
		{\sl ${}^{1}$Faculty of Physics, Shahrood University of Technology, Shahrood, Iran.}\\
		
		{\sl ${}^{2}$Department of Physics and Research Institute of Natural Science,\\
			College of Natural Science,\\
			Gyeongsang National University, Jinju 660-701, Korea.}
		
	{\sl ${}^{3}$Department of Physics, Akdeniz University, Campus 07058, Antalya, Turkey.}

    {\sl ${}^{4}$Department of Physics, University of Hradec Kr\'alov\'e,
		Rokitansk\'eho 62, 500 03 Hradec Kr\'alov\'e, Czechia.}

		\vspace{5pt}
		E-mail:  {$
			{}^{\dag}$n.farahani993@gmail.com}

		\vspace{10pt}

	\end{center}
	
	\begin{abstract}
		
		In this paper, we have used the extended generalized uncertainty principle to investigate the Unruh temperature and thermodynamic properties of a black hole. We started with a brief perusal of the Heisenberg uncertainty principle and continue with some physical and mathematical discussion for obtaining the generalized and the extended generalized uncertainty principle. Then, we obtained the Unruh temperature, mass-temperature, specific heat, and entropy functions of a black hole. We enriched the paper with graphical analysis as well as their comparisons.

	\end{abstract}
	

\section{Introduction}	
According to the de Broglie relationship, every microscopic particle exhibits both wave and particle characters. In 1927 W. Heisenberg stated that the exact position and exact momentum of microscopic particles as small as electrons could never be measured simultaneously. Later, this phenomenon was called the Heisenberg uncertainty principle (HUP) \cite{1}. Recent researches on the fields of the string theory \cite{2}, non-commutative geometries \cite{3}, black hole physics \cite{4}, and quantum gravity \cite{5} proved the existence of a minimal length. However, the HUP does not speculate a minimal length, instead, generalized uncertainty principle (GUP)  predicts a value in the scale of the Planck length \cite{6}. Alike the GUP, extended uncertainty principle (EUP) envisages a minimum measurable momentum value. Two authors of this manuscript investigated the specific heat and entropy functions of a Schwarzschild black hole with the simplest form of the EUP in \cite{7}. In an (anti)-de Sitter ((A)dS) background, the EUP yields a correction term that is proportional to the (A)dS radius \cite{8, 9, 9a}. Mignemi reported that this term could be derived from the geometry of the (A)dS spacetime as well \cite{10, 10a}. The extended generalized uncertainty principle (EGUP) is the linear combination of the GUP and EUP, and it demonstrates the high energy and large length scale modifications together \cite{11}. The EGUP is employed to obtain the Hawking temperature in the (A)dS black holes \cite {12}. Hawking temperature is a measure of black hole radiation for an observer who is in an inertial reference frame outside the black hole \cite{13, 14}. Unlike the Hawking temperature, Unruh temperature is a measure of black hole radiation observed by a uniformly accelerating detector in a vacuum field \cite{15, 16, 16a, 16b, 16c, 16d, 16e}.

In this paper, we intend to obtain a modified expression of the Unruh temperature, mass, specific heat, and entropy functions of a black hole by employing the EGUP.  We organized the paper as follows: In section \ref{sec2} we briefly introduce the HUP, GUP, EUP, and EGUP formalisms. In section \ref{sec3} we derive an expression for the Unruh temperature in the EGUP and compare it with the HUP, GUP, EUP limit ones. In section \ref{sec4},  we explore the mass-temperature, specific heat and entropy functions. We conclude the paper in section \ref{sec5}.

\section{Formalism of the uncertainty principles}\label{sec2}
In conventional quantum mechanics, the well-known HUP is written as \cite{17}
\begin{eqnarray}
  \Delta X_i \Delta P_j &\ge&\frac{\hbar \delta_{ij}}{2}.
\end{eqnarray}
In an other representation where energy and time are considered instead of the momentum and position quantites, the HUP  becomes  \cite{18, 19, 20, 20a}
\begin{eqnarray}
  \Delta E \Delta T  &\ge&  \frac{\hbar}{2}.
\end{eqnarray}
The GUP that was proposed in the context of the string theory \cite{21} and black hole Gedanken experiment \cite{21},  has the generic form of  \cite{23}
\begin{eqnarray}
 \Delta X_i \Delta P_j &\ge&  \frac{\hbar \delta_{ij}}{2}\Big[1+ \beta^2  (\Delta P_i)^2 \Big].
\end{eqnarray}
Here, $\beta^2 \equiv \frac{\beta_0}{\hbar^2} l_p^2$ where $\beta_0$ is the unit correction constant and $l_P$ is the Planck length. Unlike the GUP which is assumed to have played an important role in the early days of our universe, EUP is considered to play a role in the latter times of the universe and is written in the form \cite{24, 25, 26, 27}: \begin{eqnarray}
  \Delta X_i \Delta P_j &\ge&  \frac{\hbar \delta_{ij}}{2}\Big[1+ \alpha^2 (\Delta X_i)^2 \Big].
\end{eqnarray}
Here, $\alpha^2\equiv \frac{1}{l_H^2}$ where $l_H$ denotes the (A)dS radius. In \cite{11}, Bolen \emph{et al.} defined the EGUP
\begin{eqnarray}
  \Delta X_i \Delta P_j &\ge&  \frac{\hbar \delta_{ij}}{2}\Big[1+\alpha^2 (\Delta X_i)^2 + \beta^2(\Delta P_i)^2  \Big]. \label{EGUP}
\end{eqnarray}
to predict the temperature of the event horizon in (A)dS spacetime. In the next section, we are going to employ the EGUP to explore the Unruh temperature.

\section{Unruh temperature in the EGUP}\label{sec3}
\begin{figure}[ht]
		\centering
		\includegraphics[scale=0.58]{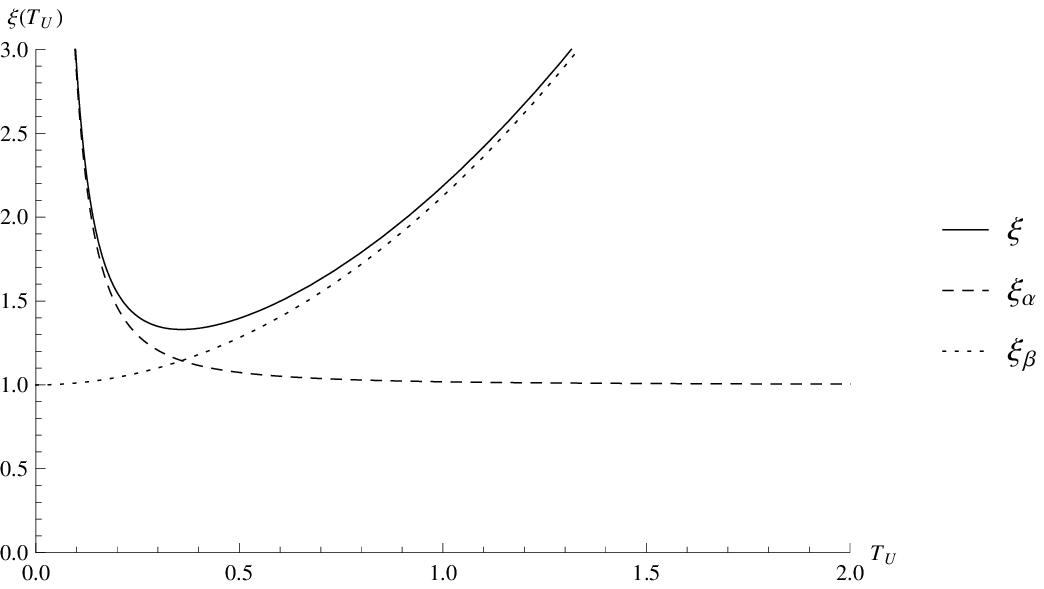}
		\caption{Plot of the modified temperature-Unruh temperature relation for the $\xi ,  \xi_{\alpha} $ and $ \xi_{\beta} $ with $\left| \Lambda \right|=0.5$  and $\beta_0 l^2_P =0.5$ where we set $\h=c=k=\pi=1$. For these values, eqs. (\ref{TUcritic}) and (\ref{etaTU}) give $T_U^C=\frac{\sqrt{2}}{3}$, and $\xi_\alpha\left(T_U^C\right)=\xi_\beta\left(T_U^C\right)=\frac{9}{8}$, respectively.} \label{fig1}
	\end{figure}

We use $\Delta E=c \Delta p$ relation in eq.~(\ref{EGUP}) and we find
\begin{eqnarray}
  \Delta X \Delta E &\ge& \frac{\h c}{2}\left(1+  \frac{|\Lambda|}{3} (\Delta X)^2 + \frac{\beta_0 l_p^2}{\hbar^2 c^2} (\Delta E)^2\right), \label{4}
\end{eqnarray}
Note that $\alpha^2\equiv \frac{|\Lambda|}{3}$, and $|\Lambda|$ is the cosmological constant \cite{10, 11, 27}. We solve the quadratic equation for $\Delta E$ and take the solution with negative sign in front of the square root because the other one does not provide a physically meaningful result \cite{26}.
\begin{eqnarray}
  \Delta E  &=& \frac{\h c}{\beta_0 l_P^2}\Delta X \left(1-\sqrt{1-\frac{\beta_0 l^2_P }{ (\Delta X)^2} - \frac{ \beta_0  l^2_P  |\Lambda|}{3}}\right).
\end{eqnarray}
We expand the square root term to its Taylor series up to second order terms and we neglected from order $O(|\Lambda|^2)$ and $O(\beta^2_0)$. We obtain
\begin{eqnarray}
  \Delta E  &=& \frac{\h c}{2\Delta X} \left[1+ \frac{ |\Lambda| (\Delta X)^2}{3}+\frac{\beta_0 l^2_P }{2}\bigg(\frac{1}{2(\Delta X)^2}+\frac{|\Lambda|}{3}\bigg) \right]. \label{EGUP2}
 \end{eqnarray}
We follow \cite{28, 29, 30}, and then, the distance along which each particle must be accelerated in order to create $N$ pairs is $\Delta X$ that we use $\Delta X= \frac{2N c^2}{a}$ and  $\Delta E=\frac{3}{2} k_B T$ in eq.~(\ref{EGUP2}). Note that  $\Delta E$ denotes the energy fluctuation during the $N$ pair production while $a$ is the acceleration of the frame. We obtain the temperature function in the EGUP as
\begin{eqnarray}
  T^{EGUP} &=& T_U\left[1 + \left| \Lambda \right| \frac{4\pi^2 c^4 }{27 a^2} + \frac{ \beta_0 l_P^2}{2}\bigg(\frac{9 a^2}{8 \pi^2 c^4} + \frac{\left| \Lambda \right| }{3}\bigg) \right]. \label{TEGUP}
\end{eqnarray}
Here, $T_U$ is the Unruh temperature \cite{16, sik}, it is written in the form of
\begin{eqnarray}
	T_U &=&\frac{\h a}{2\pi k_B c}. \label{Tunruh}
\end{eqnarray}
Alternatively, we express the derived temperature in terms of the Unruh temperature as follows:
 \begin{eqnarray}
  T^{EGUP} &=& T_U\left[1 + \frac{ \left| \Lambda \right|}{27} \left(\frac{\h c}{k_BT_U}\right)^2 + \frac{9\beta_0 l_P^2 }{4} \left(\frac{k_BT_U}{\h c}\right)^2  + \frac{\beta_0 l_P^2 \left| \Lambda \right| }{6}\right]. \label{TEGUPalt}
\end{eqnarray}
Note that, when the cosmological constant tends to zero, the Unruh temperature in the EGUP formalism reduces to the Unruh temperature in the GUP formalism. Therefore, we employ
$|\Lambda | \rightarrow 0$ in eq.~(\ref{TEGUPalt}) and we find \cite{28}
\begin{eqnarray}
  T^{GUP} &=& T_U\left[1 + \frac{9\beta_0 l_P^2 }{4} \left(\frac{k_BT_U}{\h c}\right)^2  \right], \label{TGUP}
\end{eqnarray}
This result is in agreement with \cite{7}. Alike, in the limit where the Planck length goes to zero,  $ T^{EGUP}\rightarrow T^{EUP}$.
In this case, we obtain
\begin{eqnarray}
  T^{EUP} &=& T_U\left[1 + \frac{ \left| \Lambda \right|}{27} \left(\frac{\h c}{k_BT_U}\right)^2 \right], \label{TEUPalt}
\end{eqnarray}
as found in \cite{28}. Finally, we would like to emphasize that in the HUP limit, where the cosmological constant and the Planck length vanish, we get
\begin{eqnarray}
  T^{HUP} &=& T_U. \label{THUPalt}
\end{eqnarray}
To have a better understanding of the characteristic behavior of the modified Unruh temperatures in the different formalisms, we define
\begin{eqnarray}
  \xi \equiv \frac{T^{EGUP}}{T_U}, \quad
  \xi_{\alpha} \equiv \frac{T^{EUP}}{T_U}, \quad
  \xi_{\beta} \equiv \frac{T^{GUP}}{T_U}.
\end{eqnarray}
Then, we plot $\xi$, $\xi_{\alpha}$, and $\xi_{\beta}$ versus the $T_U$ in fig. \ref{fig1}. We observe similar behaviors. In the low  Unruh temperature values $\xi$ and $\xi_{\alpha}$, and in the high Unruh temperature values $\xi$ and $\xi_{\beta}$ illustrate the same characteristic changes. However, $\xi_{\alpha}$ and $\xi_{\beta}$ do not present a similar behavior. This result confirms the agreement on the fact that the GUP and EUP have modified the early and late time dynamic of the universe. At a certain temperature
\begin{eqnarray}
  T_U^C &=& \frac{\h c}{3k_B}\sqrt{\frac{2 \alpha}{\h \beta}}. \label{TUcritic}
\end{eqnarray}
 $\xi_{\alpha}$ and $\xi_{\beta}$ have the same universal value
 \begin{eqnarray}
   \xi_{\alpha}\left(T_U^C\right) &=& \xi_{\beta}\left(T_U^C\right)=1+ \frac{\h }{2} \alpha \beta. \label{etaTU}
 \end{eqnarray}
that depends only on the cosmological constant and the Planck length.

\section{Mass-temperature, specific heat and entropy functions in the EGUP}\label{sec4}
\begin{figure}[hb]
		\centering
		\includegraphics[scale=1]{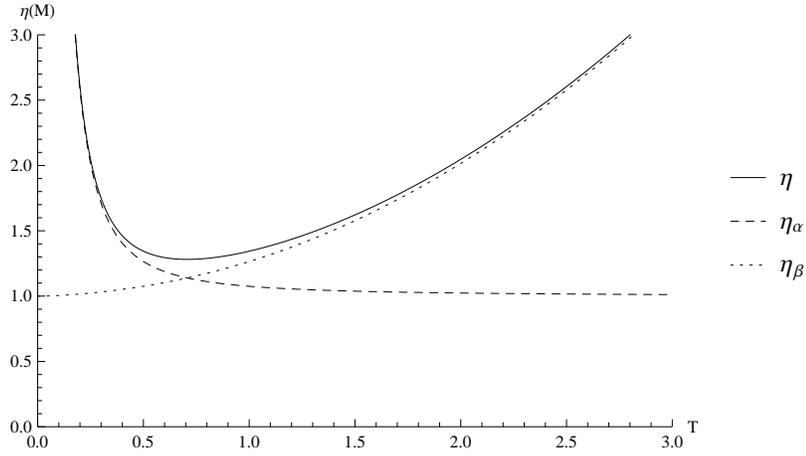}
		\caption{Plot of the mass-temperature relation for the $\eta, \eta_{\alpha}$ and $\eta_{\beta}$ with $\al=0.5$  and $\beta =0.5$ where we set $\h=c=k=1, 8 \pi G =1$. For these values,  eqs. (\ref{MTc}) and (\ref{Metac}) give $T^C=\sqrt{\frac{1}{2}}$, and $\eta_\alpha\left(T^C\right)=\eta_\beta\left(T^C\right)=\frac{41}{36}$, respectively.} \label{fig2}
	\end{figure}

\begin{figure}[ht]
		\centering
		\includegraphics[scale=1]{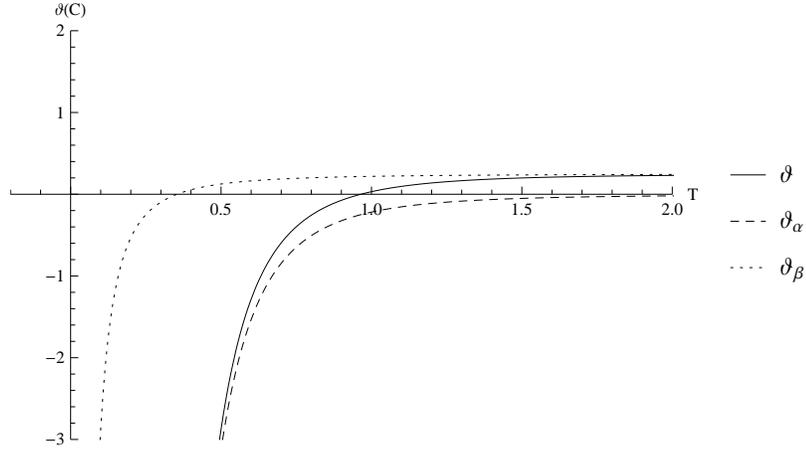}
		\caption{
			Plot of the specific heat-temperature relation for the $\vartheta$, $\vartheta_{\alpha}$ and $\vartheta_{\beta}$ with $\alpha=0.5$  and $\beta =0.5$ where we set $\h=c=k=1, 8 \pi G =1$.} \label{fig3}
	\end{figure}

\begin{figure}[hb]
	\centering
\includegraphics[scale=1]{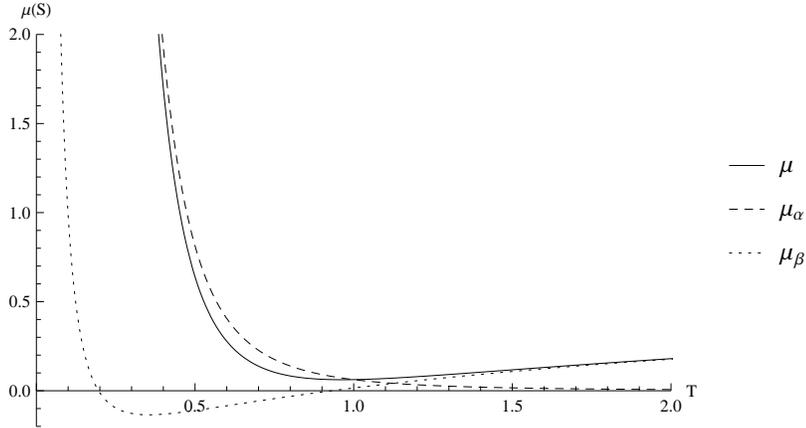}
\caption{
	Plot of the entropy-temperature relation for the $\mu$, $\mu_{\alpha}$ and $\mu_{\beta}$ with $\alpha=0.5$  and $\beta =0.5$ where we set $\h=c=k=1, 8 \pi G =1$. For these values, the critical temperature is found to be equal to $1.124$.} \label{fig4}
\end{figure}

In this section, at first, we examine the mass-temperature relation for a Schwarzschild black hole with a mass $M$ in the EGUP formalism. Then, we investigate the specific heat and entropy functions of the black hole. We assume a pair of photons near the surface. We suppose one of the photons has a negative effective energy $-E$ and it is absorbed by the black hole. Whereas, the other one has $+E$ and is emitted to an asymptotic distance from the black hole. The characteristic energy $E$ of the emitted photons may be estimated from the HUP.  Nearby the horizon of the black hole, the position uncertainty of the photon is assumed to be proportional to the Schwarzschild radius, $r_s \equiv \frac{2GM}{c^2}$, \cite{32}.
\begin{eqnarray}
   \Delta X &\equiv& \eta r_s.
\end{eqnarray}
Here, $G$ is the Newton's universal gravitational constant and $\eta$  is a scale factor. Since, photon is a massless quantum particle, its momentum uncertainty is defined with its temperature \cite{7}
\begin{eqnarray}
  \Delta P  &\equiv&  \frac{k_B T}{c}.
\end{eqnarray}
We substitute the position and momentum uncertainty expression into eq.~(\ref{EGUP}). We obtain a quadratic equation of mass and temperature in the form of
\begin{eqnarray}
  \al^2\left(\frac{2 G \eta }{c^2}\right)^2 M^2  - 2 \left(\frac{2 G \eta }{c^2}\right)\left(\frac{k_B T}{\h c}\right) M + \left[1+ (\h \beta)^2 \left(\frac{ k_B T}{\h c }\right)^2\right]  &=& 0,
\end{eqnarray}
We exclude the solution with plus sign since it has no evident physical meaning \cite{7}. Then, we express the mass-temperature function in the EGUP with
\begin{eqnarray}
M^{EGUP} &=& \frac{1}{\al^2 }\left(\frac{c^2}{2 G \eta }\right)\left(\frac{k_B T}{\h c}\right) \left[1-\sqrt{1-\left(\al \h \beta\right)^2 - \left(\al \frac{\h c}{k_B T}\right)^2}\right]. \label{MEGUP0}
\end{eqnarray}
By way of same manipulation done in sec. \ref{sec3}, we Taylor expand the square root term and present the mass-temperature equation of the Schwarzschild black hole under the EGUP formalism as
\begin{eqnarray}
M^{EGUP} &=& \frac{1}{2}\left(\frac{c^2}{2 G \eta }\right)\left(\frac{\h c}{k_B T}\right)  \left[1+\frac{\al^2}{4}\left(\frac{\h c}{k_B T}\right)^2 +(\h \beta)^2  \left(\frac{k_B T}{\h c}\right)^2 +2\left(\frac{\al \h \beta}{2}\right)^2\right].  \label{MEGUP}
\end{eqnarray}
We take $\alpha=\beta=0$ to deduce the HUP limit. We find
\begin{eqnarray}
M^{HUP} &=& \frac{1}{2}\left(\frac{c^2}{2 G \eta }\right)\left(\frac{\h c}{k_B T}\right). \label{MHUP}
\end{eqnarray}
We determine the value of the scale factor be equal to $2 \pi $ by matching eq.~(\ref{MHUP}) with the Hawking temperature \cite{7, 32}. Then, we derive the EUP limit of the mass-temperature function by employing $\beta=0$ in eq.~(\ref{MEGUP})
\begin{eqnarray}
M^{EUP} &=& \left(\frac{c^2}{8\pi G  }\right)\left(\frac{\h c}{k_B T}\right)  \left[1+\frac{\al^2}{4}\left(\frac{\h c}{k_B T}\right)^2 \right].  \label{MEUP}
\end{eqnarray}
Alike, we deduce the GUP limit of the mass-temperature function by employing $\alpha=0$ in eq.~(\ref{MEGUP})
\begin{eqnarray}
M^{GUP} &=& \left(\frac{c^2}{8\pi G  }\right)\left(\frac{\h c}{k_B T}\right)  \left[1+(\h \beta)^2  \left(\frac{k_B T}{\h c}\right)^2 \right].  \label{MGUP}
\end{eqnarray}
We find a minimum value of the temperature out of  eq.~(\ref{MEGUP0})
\begin{eqnarray}
  T &\ge& T^{EGUP}_{min} = \left(\frac{\h c }{k_B}\right)\frac{\al }{\sqrt{1-(\al \h \beta )^2}},
\end{eqnarray}	
As beta tends to zero, we obtain the minimum value of the temperature in the EUP \cite{7}
\begin{eqnarray}
  T &\ge& T^{EUP}_{min} = \left(\frac{\al\h c }{k_B}\right),
\end{eqnarray}
In the EGUP limit for $(\alpha \h \beta)^2<1$, we investigate the lowest value of the temperature, $T_0^{EGUP}$, that  minimizes the black hole mass function. We find
\begin{eqnarray}
  T_0^{EGUP}&=&\left(\frac{\h c}{k_B }\right) \sqrt{\frac{(\al \h \beta)^2+2}{(2\h \beta)^2}}
  \left[1+ \sqrt{1+\frac{12 (\alpha \h \beta)^2}{\big((\al \h \beta)^2+2\big)^2}} \right]^{1/2 }.
\end{eqnarray}
Next, we follow \cite{7} and examine the difference of mass-temperature functions in the different limits.
\begin{eqnarray}
  \Delta M^{EGUP} &\equiv& M^{EGUP}- M^{HUP}= M^{HUP} \left[\frac{\al^2}{4}\left(\frac{\h c}{k_B T}\right)^2 +(\h \beta)^2  \left(\frac{k_B T}{\h c}\right)^2 +2\left(\frac{\al \h \beta}{2}\right)^2\right] \\
  \Delta M^{EUP}  &\equiv& M^{EUP}- M^{HUP}= M^{HUP} \left[\frac{\al^2}{4}\left(\frac{\h c}{k_B T}\right)^2 \right], \\
  \Delta M^{GUP}  &\equiv& M^{GUP}- M^{HUP}= M^{HUP} \left[(\h \beta)^2  \left(\frac{k_B T}{\h c}\right)^2 \right].
\end{eqnarray}
Therefore, we obtain
\begin{eqnarray}
  \Delta M^{EGUP} &=& \Delta M^{EUP}+ \Delta M^{GUP} + M^{HUP} \frac{ \left( \al \h \beta \right)^2}{2}.
\end{eqnarray}
Before we proceed to discuss the thermal properties of the black hole, we introduce three functions,
\begin{eqnarray}
  \eta\equiv\frac{M^{EGUP}}{M^{HUP}}, \qquad \eta_{\alpha}\equiv\frac{M^{EGUP}}{M^{GUP}}, \quad \eta_{\beta}\equiv\frac{M^{EGUP}}{M^{EUP}}.
\end{eqnarray}
to illustrate the characteristic behavior of the mass-temperature in three limits. We plot $\eta$, $\eta_{\alpha}$, and $\eta_{\beta}$ versus temperature in Fig. \ref{fig2}. We observe that $\eta$ has a similar behavior  with $\eta_{\alpha}$ and $\eta_{\beta}$ at low and high temperatures, respectively. At a certain temperature value,
\begin{eqnarray}
  T^C &=& \frac{\h c}{k_B} \sqrt{\frac{\alpha}{2 \h \beta}}. \label{MTc}
\end{eqnarray}
$\eta_{\alpha}$ and $\eta_{\beta}$ become equal:
\begin{eqnarray}
  \eta_{\alpha}\left(T^C\right) &=& \eta_{\beta}\left(T^C\right) = 1+(\alpha \h \beta)- \frac{(\alpha \h \beta)}{(\alpha \h \beta)+2}. \label{Metac}
\end{eqnarray}
Next, we investigate the specific heat function of the black hole. We employ the definition $C \equiv c^2 \frac{dM}{dT}$. First, we obtain the specific heat in the HUP limit as
\begin{eqnarray}
  C^{HUP} &=& -\frac{c^2}{T}M^{HUP}. \label{CHUP}
\end{eqnarray}
Then, we use eq. (\ref{MEGUP}) to derive the specific heat function in the EGUP limit.
\begin{eqnarray}
  C^{EGUP} &=& C^{HUP} \left[1+\frac{3\al^2}{4}\left(\frac{\h c}{k_B T}\right)^2 -(\h \beta)^2  \left(\frac{k_B T}{\h c}\right)^2 +2\left(\frac{\al \h \beta}{2}\right)^2\right].  \label{CEGUP}
\end{eqnarray}
We employ  $\beta \rightarrow 0$ and $\alpha \rightarrow 0$ conditions in eq.~(\ref{CEGUP})  to deduce the specific function in the EUP and GUP limits. We find
\begin{eqnarray}
  C^{EUP} &=& C^{HUP} \left[1+\frac{3\al^2}{4}\left(\frac{\h c}{k_B T}\right)^2 \right], \label{CEUP} \\
  C^{GUP} &=& C^{HUP} \left[1 -(\h \beta)^2  \left(\frac{k_B T}{\h c}\right)^2 \right].  \label{CGUP}
\end{eqnarray}
Next, we define
\begin{eqnarray}
  \Delta C^{EGUP} &\equiv& C^{EGUP}- C^{HUP}= C^{HUP} \left[\frac{3\al^2}{4}\left(\frac{\h c}{k_B T}\right)^2 -(\h \beta)^2  \left(\frac{k_B T}{\h c}\right)^2 +2\left(\frac{\al \h \beta}{2}\right)^2\right], \\
  \Delta C^{EUP}  &\equiv& C^{EUP}- C^{HUP}= C^{HUP} \left[\frac{3 \al^2}{4}\left(\frac{\h c}{k_B T}\right)^2 \right], \\
  \Delta C^{GUP}  &\equiv& C^{GUP}- C^{HUP}= C^{HUP} \left[-(\h \beta)^2  \left(\frac{k_B T}{\h c}\right)^2 \right].
\end{eqnarray}
Similar to the analysis that is carried on the mass temperature function, we observe that the contributions to the specific function can be expressed separately from each uncertainty limit.
\begin{eqnarray}
  \Delta C^{EGUP} &=& \Delta C^{EUP}+ \Delta C^{GUP} + C^{HUP} \frac{ \left( \al \h \beta \right)^2}{2}.
\end{eqnarray}
Before we proceed to examine the entropy function of the black hole, we introduce the following functions
\begin{eqnarray}
  \vartheta \equiv C^{EGUP}- C^{HUP}, \qquad \vartheta_{\alpha}\equiv C^{EGUP}- C^{GUP} , \quad \vartheta_{\beta}\equiv C^{EGUP}- C^{EUP}.
\end{eqnarray}
In fig. \ref{fig3} we present their behaviors versus the temperature. We observe that at low temperatures $\vartheta_{\beta}$ and $\vartheta$, at high temperatures $\vartheta_{\alpha}$ and $\vartheta$ have similar shapes. We point out that $\vartheta_{\alpha}$ is always in the negative value region.
\\
Finally we derive the entropy functions via the definitions given in \cite{7, 33, 34}.  We find
\begin{eqnarray}
   S^{HUP}  &=& \frac{c^2}{2T}M^{HUP}, \label{SHUP} \\
   S^{EGUP} &=& S^{HUP} \left[1+\frac{3\al^2}{8}\left(\frac{\h c}{k_B T}\right)^2 +(\h \beta)^2  \left(\frac{k_B T}{\h c}\right)^2 \ln T +2\left(\frac{\al \h \beta}{2}\right)^2\right],  \label{SEGUP}\\
   S^{EUP} &=& S^{HUP} \left[1+\frac{3\al^2}{8}\left(\frac{\h c}{k_B T}\right)^2 \right], \label{SEUP} \\
   S^{GUP}  &=& S^{HUP} \left[1 + 2 (\h \beta)^2  \left(\frac{k_B T}{\h c}\right)^2 \ln T \right].  \label{SGUP}
\end{eqnarray}
Next, we define variation functions to illustrate the contributions of each uncertainty limit to the total entropy function.
\begin{eqnarray}
  \Delta S^{EGUP} &\equiv& S^{EGUP}- S^{HUP}= S^{HUP} \left[\frac{3\al^2}{8}\left(\frac{\h c}{k_B T}\right)^2 +(\h \beta)^2  \left(\frac{k_B T}{\h c}\right)^2\ln T +2\left(\frac{\al \h \beta}{2}\right)^2\right], \\
  \Delta S^{EUP}  &\equiv& S^{EUP}- S^{HUP}= S^{HUP} \left[\frac{3 \al^2}{8}\left(\frac{\h c}{k_B T}\right)^2 \right] \\
  \Delta S^{GUP}  &\equiv& S^{GUP}- S^{HUP}= S^{HUP} \left[(\h \beta)^2  \left(\frac{k_B T}{\h c}\right)^2 \ln T \right].
\end{eqnarray}
Alike the mass-temperature and specific heat functions, these variation functions also satisfy
\begin{eqnarray}
  \Delta S^{EGUP} &=& \Delta S^{EUP}+ \Delta S^{GUP} + S^{HUP} \frac{ \left( \al \h \beta \right)^2}{2}.
\end{eqnarray}
We consider the following functions
\begin{eqnarray}
  \mu\equiv S^{EGUP}- S^{HUP}, \qquad \mu_{\alpha}\equiv S^{EGUP}- S^{GUP} , \quad \mu_{\beta}\equiv S^{EGUP}- S^{EUP}.
\end{eqnarray}
Then, we present the entropy function behaviors in fig. (\ref{fig4}). We observe that $\mu$ mimics the $\mu_\alpha$ and  $\mu_\beta$ functions at low and high temperature values, respectively. $\mu_\alpha$ and $\mu_\beta$ has the same value at a critic temperature which can be evaluated from the root of the following function.
\begin{eqnarray}
  \left(\frac{k_B T}{\h c}\right)^4 \ln T &=& \frac{3}{8} \left(\frac{\alpha}{\h \beta}\right)^2. \label{STcritic}
\end{eqnarray}

\section{Conclusion}\label{sec5}
In this letter, we examined the Unruh temperature and the thermodynamic properties of a black hole by employing the extended generalized uncertainty principle which is a summation of generalized and extended uncertainty principles. We showed that the modified Unruh temperature of the EGUP limit has similar behavior with the GUP and EUP limits in high and low-temperature values, respectively.  Our investigation on a black hole's mass-temperature function in the EGUP limit ended with an expression that describes the minimum temperature value. Moreover, the relatively defined mass functions in the different limits yielded similar behaviors at low and high temperatures. Furthermore, we investigated the specific heat and the entropy functions of the black hole in the EGUP formalism. Alike to the mass-temperature functions, we found that the contributions of the EGUP formalism can separately be expressed in the specific heat and entropy functions. The relatively defined functions of specific heat and entropy showed that at low-temperature and high-temperature EGUP functions behave as EUP and GUP functions.

\section{Acknowledgment}
The authors thank the referees for a thorough reading of our manuscript and for constructive suggestion. One of the author, B.C. L\" utf\"uo\u{g}lu, was partially supported by the Turkish Science and Research Council (T\"{U}B\.{I}TAK).


\newpage

\end{document}